\documentclass[lettersize,journal]{IEEEtran}
\usepackage{amsmath,amsfonts,amssymb}
\usepackage{tcolorbox}
\usepackage{tikz}
\usetikzlibrary{positioning}
\usepackage{pifont}
\usepackage{algorithmic}
\usepackage{algorithm}
\usepackage{amsthm}

\usepackage{array}
\usepackage[caption=false,font=normalsize,labelfont=sf,textfont=sf]{subfig}
\usepackage{textcomp}
\usepackage{stfloats}
\usepackage{url}
\usepackage{verbatim}
\usepackage{graphicx}
\usepackage{cite}
\usepackage{booktabs}
\usepackage{tabularx}
\usepackage{multirow}
\usepackage{pifont}  
\usepackage{threeparttable}  
\usepackage{float}
\floatstyle{plain}
\newfloat{protocolfloat}{tbp}{lop}
\floatname{protocolfloat}{Protocol}
\newcommand{\cmark}{\ding{51}}  
\newcommand{\xmark}{\ding{55}}  
\usepackage[colorlinks=true, linkcolor=blue, citecolor=blue]{hyperref}
\newenvironment{protocol}[2][t]{%
  \begin{protocolfloat}[#1]
  \begin{tcolorbox}[
    colback=white,
    colframe=black,
    fonttitle=\bfseries,
    title=#2,
    arc=3mm,
    boxrule=1pt,
    left=5pt,
    right=5pt,
    top=5pt,
    bottom=5pt
  ]
}{%
  \end{tcolorbox}
  \end{protocolfloat}
}

\begin{document}

\title{Efficient Multi-Party Secure Comparison over Different \\Domains with Preprocessing Assistance}



\author{Kaiwen Wang, Xiaolin Chang, Yuehan Dong, Ruichen Zhang
}

\markboth{Journal of \LaTeX\ Class Files,~Vol.~14, No.~8, August~2021}%
{Shell \MakeLowercase{\textit{et al.}}: A Sample Article Using IEEEtran.cls for IEEE Journals}


\maketitle

\begin{abstract}
Secure comparison is a fundamental primitive in multi-party computation,
supporting privacy-preserving applications such as machine learning and data
analytics. 
A critical performance bottleneck in comparison protocols is their preprocessing phase, primarily due to the high cost of generating the necessary correlated randomness. Recent frameworks introduce a passive, non-colluding dealer to accelerate
preprocessing. However, two key issues still remain. First, existing dealer-assisted approaches
treat the dealer as a drop-in replacement for conventional preprocessing without
redesigning the comparison protocol to optimize the online phase. Second, most
protocols are specialized for particular algebraic domains, adversary models, or
party configurations, lacking broad generality. 
In this work, we present the first dealer-assisted $n$-party  LTBits (Less-Than-Bits) and MSB
(Most Significant Bit) extraction
protocols over both $\mathbb{F}_p$ and $\mathbb{Z}_{2^k}$, achieving perfect
security at the protocol level. By fully exploiting the dealer's capability to
generate rich correlated randomness, our $\mathbb{F}_p$ construction achieves
constant-round online complexity and our $\mathbb{Z}_{2^k}$ construction
achieves $O(\log_n k)$ rounds with tunable branching factor. All protocols are
formulated as black-box constructions via an extended ABB model, ensuring
portability across MPC backends and adversary models. Experimental results
demonstrate $1.79\times$ to $19.4\times$ speedups over state-of-the-art MPC frameworks,
highlighting the practicality of our protocols for comparison-intensive MPC
applications.
\end{abstract}

\begin{IEEEkeywords}
Multi-party computation, secure comparison, dealer-assisted preprocessing, 
perfect security.
\end{IEEEkeywords}

\section{Introduction}
\label{Sec:introduction}
Secure multi-party computation (MPC)~\cite{yao1982protocols} enables multiple parties to jointly compute functions over their private inputs without revealing anything beyond the output. In practice, such functions often involve complex compositions of multiple operations, while modern MPC frameworks~\cite{dohmen2025seec}~\cite{smajlovic2025shechi}~\cite{keller2025scale} are highly optimized primarily for arithmetic operations like addition and multiplication. However, many practical applications such as privacy-preserving machine learning~\cite{mohassel2017secureml}, secure auctions~\cite{sharma2025blockbid}, and data analytics~\cite{van2026mozaik} fundamentally require comparison operations in addition to arithmetic operations.

At a high level, a comparison operation determines the relative order of two secret values, such as whether one value is smaller than another or whether a value is negative. Although conceptually simple, secure comparison is notoriously more challenging than arithmetic operations in MPC. Arithmetic operations can be efficiently supported based on secret sharing and well-established Beaver triple techniques~\cite{beaver1991efficient}, whereas comparison does not admit such a direct formulation. MPC-based comparison protocols typically require non-trivial transformations, including bit decomposition of arithmetic shares, prefix computations over secret bits, or masked zero-testing procedures. These operations significantly increase communication overhead, interaction rounds, and preprocessing overhead compared to purely arithmetic computations, making secure comparison one of the main performance bottlenecks~\cite{diaa2024fast} in practical MPC systems. In particular, the preprocessing phase of comparison protocols often dominates overall system performance, as it involves generating complex correlated randomness such as daBits, edaBits~\cite{escudero2020improved} and function secret sharing (FSS)~\cite{boyle2015function}. Such preprocessing overhead can be several or even dozens of times higher than online costs, as observed even in state-of-the-art frameworks such as \emph{Rabbit}~\cite{makri2021rabbit}, severely limiting the practical deployment of MPC-based comparison.

To alleviate the costly preprocessing bottleneck, recent MPC frameworks introduced a passive, non-colluding dealer to accelerate the generation of correlated randomness~\cite{kei2025shaft}~\cite{jawalkar2024orca}~\cite{gupta2023sigma}~\cite{zbudila2025sok}~\cite{hastings2019sok}. In this paradigm, a dealer that does not participate in the online computation generates and distributes preprocessing data to all parties in a preprocessing phase, bypassing the expensive cryptographic procedures~\cite{sun2025sok} otherwise required. Preprocessing data exist mainly in the form of correlated randomness. While this dealer-assisted approach has shown promise in reducing preprocessing costs for general MPC tasks, its application to secure comparison still leaves significant room for improvement. In particular, we identify the following two key issues that remain unaddressed.

\textbf{Issue 1: Underutilized Dealer Capability for Online
Phase Optimization.}
Existing dealer-assisted MPC frameworks ~\cite{kei2025shaft}~\cite{jawalkar2024orca}~\cite{gupta2023sigma}~\cite{zbudila2025sok}~\cite{hastings2019sok} largely employ the dealer as a drop-in replacement for conventional preprocessing methods, generating the same types of correlated randomness such as multiplication triples and edaBits without rethinking the structure of comparison protocols. However, a passive dealer possesses a unique capability: it can efficiently generate richer and more complex forms of correlated randomness that would be prohibitively expensive to produce using standard MPC techniques. These include powers of random masks, perfectly consistent arithmetic and bitwise decompositions, and exponentially-sized preprocessing data for multi-input gates. Whether such enhanced preprocessing can enable fundamental redesigns of comparison protocols with substantially lower online round complexity and communication cost remains an open question.


\textbf{Issue 2: Lack of Generality across Algebraic Do-mains,
Adversary Models, and Party Conffgurations.}
Here, \emph{algebraic domains} refer to the algebraic structures over which computation 
is performed such as prime fields $\mathbb{F}_p$ and integer rings 
$\mathbb{Z}_{2^k}$; \emph{adversary models} specify the corruption power of 
the adversary, including whether it is \emph{passive (semi-honest)} or \emph{active (malicious)}; and \emph{party configurations} encompass the number of 
participating parties and the corruption threshold such as \emph{honest-majority} 
and \emph{dishonest-majority}. Section~\ref{subsec:Security Model} offers a detailed introduction. Most existing comparison protocols~\cite{kei2025shaft}~\cite{jawalkar2024orca}~\cite{gupta2023sigma}~\cite{damgaard2006unconditionally}~\cite{nishide2007multiparty}~\cite{reistad2009multiparty}~\cite{catrina2010improved}~\cite{lipmaa2013secure}~\cite{damgaard2019new}~\cite{duan2021acco}~\cite{cheng2025mosformer}, whether dealer-assisted or not, are designed as specialized solutions that exploit techniques available only under restricted configurations. For instance, FSS~\cite{boyle2015function} and Replicated secret sharing ~\cite{baccarini2023multi} enable highly efficient comparison~\cite{cheng2025mosformer} in the two/three-party setting but cannot be extended to the general $n$-party case. Truncation-based optimizations~\cite{damgaard2019new} are effective over rings $\mathbb{Z}_{2^k}$ but do not carry over to prime fields $\mathbb{F}_p$. These specialized techniques achieve strong performance precisely because they leverage structural properties of specific configurations. In the general $n$-party setting with arbitrary adversary models and multiple algebraic domains, such techniques are no longer applicable, and protocol designers must rely on more general primitives that typically incur higher overhead. Designing comparison protocols that simultaneously support general $n$-party computation, multiple algebraic domains, and diverse adversary models while maintaining a black-box relationship with the underlying MPC backend remains a fundamental challenge.

\begin{figure}[t]
\centering
\includegraphics[width=\columnwidth]{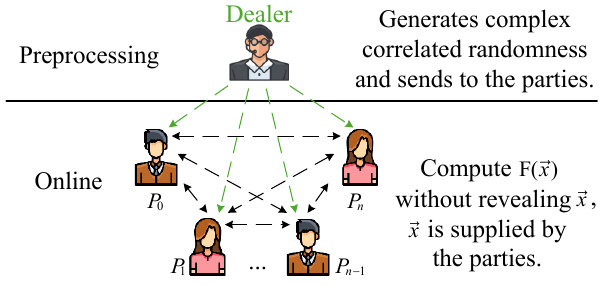}
\caption{The dealer-assisted MPC framework.}
\label{fig:system-model}
\end{figure}

Motivated by the above discussions, this apper revisits MPC-based comparison from a dealer-assisted perspective, aiming to fully exploit the dealer's preprocessing capability to redesign comparison protocols with reduced online complexity and broad generality. Figure~\ref{fig:system-model} illustrates the dealer-assisted MPC framework adopted in this work. The main contributions are summarized as follows.

\begin{itemize}
  \item \textbf{Efficient dealer-assisted comparison protocols with perfect security.}
  For \textbf{Issue 1}, we design the first dealer-assisted $n$-party LTBits (Less-Than-Bits) and MSB (Most Significant Bit) extraction protocols over
  both $\mathbb{F}_p$ and $\mathbb{Z}_{2^k}$, achieving perfect security at the protocol level. Our protocols fully leverage the dealer's efficient preprocessing capability
  to minimize online overhead:
  \begin{itemize}
    \item For $\mathbb{F}_p$, we develop a polynomial-based construction where
    the dealer generates perfectly consistent arithmetic and bitwise decomposition
    of random masks during preprocessing, eliminating statistical errors from
    edaBits-style conversions and achieving constant-round (2 rounds) online complexity.

    \item For $\mathbb{Z}_{2^k}$, we propose a multi-input AND gate optimization
    technique that utilizes the dealer to efficiently generate exponentially-sized
    preprocessing data ($2^m-1$ secret shares for $m$-input gates), achieving $O(\log_n k)$ online round complexity with
    tunable branching factor $n$.
  \end{itemize}

  \item \textbf{Black-box construction and domain independence.} For \textbf{Issue 2}, our protocols
  are formulated entirely in terms of abstract MPC functionalities via an extended
  ABB model, supporting computation under both honest/dishonest-majority and passive/active adversary models. This makes them portable across different MPC backends and algebraic domains without relying on backend-specific techniques.
\end{itemize}

We implement our protocols and conduct extensive experimental evaluations under realistic network settings (LAN/WAN), demonstrating substantial performance improvements over state-of-the-art frameworks: up to $19.4\times$ speedup for
\hyperref[protocol:msb-p]  {$\Pi_{\text{MSB}_p}$} in WAN settings with small batch sizes, and $1.79\times$ to $2.78\times$ speedup for \text{\hyperref[protocol:msb-2k]{$\Pi_{\text{MSB}_{2^k}}$}} across various
configurations compared to Rabbit.

The rest of this paper is organized as follows. Section~\ref{Sec:related work} 
reviews related work. Section~\ref{Sec:preliminaries} introduces background 
and the extended ABB model. Sections~\ref{Sec:ltbits-fp} and \ref{Sec:ltbits-f2} 
present our LTBits protocols for $\mathbb{F}_p$ and $\mathbb{Z}_{2^k}$ 
respectively. Section~\ref{Sec:msb} derives MSB extraction protocols and 
discusses perfect security. Section~\ref{sec:evaluation} presents 
experimental evaluations. Section~\ref{Sec:conclusion} concludes the paper. Due to page limitations, an incorrect MSB construction from prior work is analyzed Section I of the supplementary file, and the proof of ABB extension with multi-input AND gates is given in Section II.


\section{Related Work}
\label{Sec:related work}
The past two decades saw a large body of work on studying MPC-based comparison protocols under different adversary models, algebraic domains, and efficiency goals. This section reviews representative protocols that support the general $n$-party setting. Table~\ref{tab:related_work} summarizes these works along several key dimensions. We first discuss prior protocols and then provide a more detailed discussion of Rabbit~\cite{makri2021rabbit}, which is widely regarded as the state-of-the-art comparison framework.

\paragraph{MPC comparison protocols prior to Rabbit.}
Early work by Damg{\aa}rd \emph{et al.}~\cite{damgaard2006unconditionally}
initiated the study of secure comparison in MPC, focusing on unconditional or
information-theoretic security under honest-majority assumptions. While these
protocols established important theoretical foundations, they typically incur
a large number of communication rounds and substantial online overhead.
Nishide and Ohta~\cite{nishide2007multiparty} proposed comparison protocols over $\mathbb{F}_p$ without explicit bit-decomposition, reducing computational cost in the passive, honest-majority setting. Reistad~\cite{reistad2009multiparty} further optimized round complexity and demonstrated that constant-round comparison is achievable under certain assumptions, at the cost of additional preprocessing or relaxed adversary models.

Subsequent work by Catrina and de Hoogh~\cite{catrina2010improved} identified
comparison and truncation as major bottlenecks in integer MPC and proposed
improved primitives to enhance practical performance. Lipmaa and
Toft~\cite{lipmaa2013secure} focused on reducing online communication complexity,
achieving logarithmic or constant online costs for comparison under active
security with honest majority. More recent efforts extended comparison to
ring-based computation over $\mathbb{Z}_{2^k}$, which is particularly relevant
for privacy-preserving machine learning. Damg{\aa}rd \emph{et
al.}~\cite{damgaard2019new} studied comparison in malicious settings over
$\mathbb{Z}_{2^k}$, while Duan \emph{et al.}~\cite{duan2021acco} incorporated
comparison as a core primitive in an MPC framework supporting arithmetic
circuits. Despite these advances, most existing protocols either rely on
non-trivial preprocessing with high complexity or achieve limited reductions in
online rounds under strong adversary models.

\paragraph{Rabbit comparison framework.}
Rabbit by Makri \emph{et al.}~\cite{makri2021rabbit} represents a major milestone
in MPC-based comparison protocols. It introduces a unified framework for MPC-based comparison, in which
Less-Than-Bits (LTBits) serves as a core primitive from which Less-Than-Constant (LTC), Less-Than-Secret (LTS), and Most Significant Bit (MSB) extraction protocols are systematically derived. Rabbit~\cite{makri2021rabbit}
supports general $n$-party computation under both honest-majority and
dishonest-majority settings, and applies to multiple algebraic domains,
including $\mathbb{F}_p$ and $\mathbb{Z}_{2^k}$.

Unlike prior designs that rely on slack or bounded-input assumptions, Rabbit~\cite{makri2021rabbit} provides comparison protocols whose correctness does not depend on inflating the representation space beyond the length of the input bit. 
This allows comparison to be integrated naturally into arithmetic MPC engines over both fields and rings. As a result, Rabbit establishes a widely adopted baseline for modern MPC comparison protocols.



\begin{table*}[t]
\centering
\caption{Comparison of Related MPC Comparison Protocols}
\label{tab:related_work}
\begin{threeparttable}
\small
\begin{tabular}{lccccccc}
\toprule
\textbf{Protocol} & \textbf{Dealer} & \textbf{Black-} & \textbf{Domain} & \textbf{Security} & \textbf{Adversary/} & \textbf{Online} & \textbf{Online} \\
 & \textbf{Assisted} & \textbf{Box$^\dagger$} &  & \textbf{Level$^\ddagger$} & \textbf{Majority} & \textbf{Rounds} & \textbf{Comm.} \\
\midrule
Damg{\aa}rd et al.~\cite{damgaard2006unconditionally} 2006 & \xmark & \cmark & $\mathbb{F}_p$ & Perfect & Active/Honest & 44  & $O(k\log k)$ \\
Nishide et al.~\cite{nishide2007multiparty} 2007 & \xmark & \xmark & $\mathbb{F}_p$ & Perfect & Passive/Honest & 15 & $O(k)$ \\
Reistad~\cite{reistad2009multiparty} 2009 & \xmark & \cmark & $\mathbb{F}_p$ & Perfect & Unrestricted & 5 & $O(k)$ \\
Catrina et al.~\cite{catrina2010improved} 2010 & \xmark & \xmark & $\mathbb{F}_p$ & Statistical & Passive/Honest & 3 & $O(k)$ \\
Lipmaa et al.~\cite{lipmaa2013secure} 2013 & \xmark & \cmark & $\mathbb{F}_p$ & Perfect & Active/Honest & $O(\log k)$ & $O(\log k)$ \\
Damg{\aa}rd et al.~\cite{damgaard2019new} 2019 & \xmark & \cmark & $\mathbb{Z}_{2^k}$ & Statistical & Active/Dishonest & $O(\log k)$ & $O(k\log k)$ \\
Duan et al.~\cite{duan2021acco} 2021 & \xmark & \xmark & $\mathbb{F}_p$ & Perfect & Active/Honest & $O(\log k)$ & $O(k)$ \\
Rabbit~\cite{makri2021rabbit} 2021 & \xmark & \cmark & $\mathbb{Z}_{2^k}, \mathbb{F}_p$ & Perfect/Statistical$^*$ & Unrestricted & $O(\log k)$ & $O(k\log k)$ \\
\midrule
\textbf{Our Work (\hyperref[protocol:msb-p]{$\Pi_{\text{MSB}_p}$})} & \cmark & \cmark & $\mathbb{F}_p$ & Perfect & Unrestricted & \textbf{2} & $O(k)$ \\
\textbf{Our Work (\text{\hyperref[protocol:msb-2k]{$\Pi_{\text{MSB}_{2^k}}$}})} & \cmark & \cmark & $\mathbb{Z}_{2^k}$ & Perfect & Unrestricted & $O(\log_n k)$ & $O(k\log_n k)$ \\
\bottomrule
\end{tabular}
\begin{tablenotes}
\footnotesize
\item $^\dagger$: A protocol is marked as black-box if it is specified
solely in terms of abstract MPC functionalities and does not rely on properties
of a particular secret-sharing backend. The black-box abstraction
used in this work is formally defined and discussed in
Section~\ref{subsec:Security Model}.
\item $^\ddagger$: Security levels refer to the protocol-level security. Concrete instantiations may achieve different security levels depending on the underlying MPC backend. For example, instantiations using SPDZ$_{2^k}$ or TinyOT may provide statistical security even when the protocol achieves perfect security.
\item $k$: bit length of plaintext values; $n$: maximum branching factor for multi-input AND gates.
\item $^*$ Perfect security for $\mathbb{Z}_{2^k}$, statistical security for $\mathbb{F}_p$.

\end{tablenotes}
\end{threeparttable}
\end{table*}

\section{Preliminaries}
\label{Sec:preliminaries}

In this section, we provide the necessary background. Section~\ref{subsec:Security Model} introduces the security model and black-box design philosophy. Section~\ref{subsec:Abstract Sharing} defines abstract sharing semantics. Section~\ref{subsec:instantiation} presents concrete instantiation examples for performance evaluation. Section~\ref{protocol:ltbits2-n} reviews Rabbit's LTBits protocol~\cite{makri2021rabbit}, which serves as a foundation for our work.

\subsection{Security Model and Black-box Abstraction}
\label{subsec:Security Model}

We consider a universal multi-party computation (MPC) setting with $n$ parties
jointly evaluating a comparison functionality over secret inputs. The adversary
may be passive (semi-honest) or active(malicious) and honest-majority or dishonest-majority. 

A passive adversary follows the prescribed protocol but attempts to learn additional information from its view. An active adversary may arbitrarily deviate from the protocol; if detected, execution is immediately aborted. For a system with $n$ parties, the number of corrupted parties is bounded by $t$ such that $2t < n$ in the honest-majority setting. In contrast, in the dishonest-majority setting, the adversary may corrupt up to $n-1$ parties.

Rather than committing to a specific corruption threshold or MPC backend at the protocol-design level, our goal is to design comparison protocols that are described as \emph{black-box constructions} with respect to the underlying MPC functionality.

\subsubsection{Black-box abstraction via the ABB model}

We formalize our black-box design philosophy using an extended version of Arithmetic Black-Box (ABB) model\cite{escudero2020improved} which is an ideal
functionality in the universal composability framework\cite{canetti2001universally}. The ABB model abstracts MPC as an ideal functionality that provides fixed interfaces for operations on shared 
values. In our protocol descriptions, all steps are expressed solely through 
the abstract operations provided by ABB. Figure~\ref{fig:abb} presents the 
ideal functionalities defined in the extended ABB model. We note that our work extends the standard ABB model again by incorporating multi-input multiplication gates into the ABB interface. The correct proof of this extension follows from combining $\mathcal{F}_{\text{LinComb}}$, $\mathcal{F}_{\text{Mult}}$ and $\mathcal{F}_{\text{Open}}$ with multiplication tuples, as detailed in Section II of supplementary file. Consequently, our protocols make a purely black-box use of the underlying ABB functionality. Any backend securely realizing the ABB interface can instantiate our protocols, and the end-to-end security will naturally inherit the concrete guarantees provided by the chosen backend. 

As a consequence, we do not provide standalone security proofs for each protocol. All protocols inherit their security directly from the underlying ABB model via the composition theorem~\cite{canetti2001universally}: once a concrete backend securely realizes the ABB interface, the security of our protocols follows immediately.

\begin{figure}[t]
\centering
\begin{tikzpicture}
\node[rectangle, draw=black, fill=white, rounded corners=3mm, 
      text width=\columnwidth-8pt, inner sep=5pt, line width=1.5pt,
      align=justify] (content) at (0,0) {  
\small

\textbf{$\mathcal{F}_{\text{Input}}$:} On input (Input, $P_i$, type, id, $x$) from $P_i$ and (Input, $P_i$, type, id) from all other parties, with id a fresh identifier, type $\in$ \{binary, arithmetic\} and $x \in M$, store (type, id, $x$).

\smallskip
\textbf{$\mathcal{F}_{\text{LinComb}}$:} On input (LinComb, type, id, $(\mathrm{id}_j)_{j=1}^m$, type, $c$, $(c_j)_{j=1}^m$), where each $\mathrm{id}_j$ is stored in memory, compute $y = c + \sum_j x_j \cdot c_j$ modulo 2 if type = binary and modulo $M$ if type = arithmetic, and store (type, id, $y$).

\smallskip
\textbf{$\mathcal{F}_{\text{Mult}}$:} On input (Mult, type, id, $\mathrm{id}_1$, $\mathrm{id}_2$) from all parties, retrieve (type, $\mathrm{id}_1$, $x$), (type, $\mathrm{id}_2$, $y$), compute $z = x \cdot y$ modulo 2 if type = binary and modulo $M$ if type = arithmetic, and store (id, $z$).

\smallskip
\textbf{$\mathcal{F}_{\text{m-Mult}}$:} On input (m-Mult, type, id, $(\mathrm{id}_j)_{j=1}^m$) from all parties, retrieve ((type, $\mathrm{id}_1$, $x_1$), $\ldots$, (type, $\mathrm{id}_m$, $x_m$)), compute $z = \prod_{j=1}^m x_j$ modulo 2 if type = binary and modulo $M$ if type = arithmetic, and store (id, $z$). This extends standard multiplication to support m-input AND gates.

\smallskip
\textbf{$\mathcal{F}_{\text{B2A}}$:} On input (ConvertB2A, id, id$^\prime$) from all parties, retrieve (binary, id$^\prime$, $x$) and store (arithmetic, id, $x$).

\smallskip
\textbf{$\mathcal{F}_{\text{Open}}$:} On input (Output, type, id) from all honest parties, retrieve (type, id, $y$) and output it to the adversary. Wait for an input from the adversary; if this is \texttt{Proceed} then output $y$ to all parties, otherwise output \texttt{Abort}.

\smallskip
\textbf{$\mathcal{F}_{\text{daBits}}$:} On input (daBits, $k$, id, id$^\prime$) from all parties, sample $r_1, \ldots, r_k \xleftarrow{\$} \mathbb{F}_2$ uniformly at random and store (binary, id$^\prime$, $r_i$) and (arithmetic, id, $r_i$) for $i \in [1, k]$.

\smallskip
\textbf{$\mathcal{F}_{\text{edaBits}}$:} On input (edaBits, $m$, $k$, id, id$^\prime$) from all parties, for $j \in [1, k]$, sample $r_j \xleftarrow{\$} \mathbb{Z}_{2^m}$ uniformly at random, compute bit decomposition $r_j = \sum_{i=0}^{m-1} r_{j,i} \cdot 2^i$ where $r_{j,i} \in \{0, 1\}$, and store (arithmetic, id, $r_j$), (binary, id$^\prime$, $r_{j,i}$) for $i \in [0, m-1]$, $j \in [1,k]$.
};

\node[rectangle, draw=black, fill=lightgray, rounded corners=2mm, 
      inner sep=3pt, line width=1pt, anchor=west] (title) 
      at ([xshift=5mm]content.north west) 
      {\textbf{Functionality} $\mathcal{F}_{\mathrm{ABB}}^{+}$};

\end{tikzpicture}
\caption{Ideal functionality for the MPC arithmetic black-box model modulo $M$, where $M$ is either a ring or a finite field.}
\label{fig:abb}
\end{figure}

\subsubsection{Introduction of dealer and design rationale}

In our concrete protocol constructions, we adopt a \textit{preprocessing-and-online} paradigm. The preprocessing phase is performed by a passive, non-colluding dealer that generates and distributes the appropriate preprocessing data 
to all parties. This process does not require participation from the computing 
parties and therefore does not affect the security of the online phase. 

The dealer-based preprocessing approach enables efficient generation of complex 
correlated randomness, such as powers of random masks or multiplication tuples, 
that would be expensive to produce using standard MPC techniques. The dealer 
only participates in the preprocessing phase and learns no information about the 
actual inputs processed during the online computation.

\subsubsection{Instantiation for evaluation}

While the protocol design is black-box, concrete instantiations are
necessary to implement the protocols and evaluate their practical performance.
For this purpose, we instantiate our protocols using two representative MPC
backends in Section~\ref{subsec:instantiation}: an SPDZ-style authenticated
arithmetic sharing backend~\cite{keller2020mp}~\cite{cramer2018spd} for computations over $\mathbb{F}_p$ and
$\mathbb{Z}_{2^k}$, and a TinyOT-style sharing backend~\cite{burra2021high} for Boolean
circuits. These instantiations are used only to demonstrate feasibility and
measure performance; they do not limit the applicability of our protocols to
other MPC frameworks that support the same abstract operations.

\subsection{Abstract Sharing Semantics and Notation}
\label{subsec:Abstract Sharing}

\begin{table}[t]
\centering
\caption{Notation Used Throughout the Paper}
\label{tab:notation}
\begin{tabularx}{\columnwidth}{lX}
\toprule
\textbf{Notation} & \textbf{Description} \\
\midrule
$\mathbb{Z}_{2^k}$ & The ring of integers modulo $2^k$, where $k$ is a positive integer. \\
$\mathbb{F}_N$ & Finite field of order $N$, where $N \in \{ p,2,{2^s}\}$, $p$ is a large prime, $s$ is a positive integer. \\
$\vec{x}, \vec{x}[i]$ & Vector $\vec{x}$ and the $i$-th element of vector $\vec{x}$. \\
$x[i]$ & The $i$-th bit of value $x$. \\
$1\{x < y\}$ & Indicator function: return 1 if $x < y$, otherwise return 0. \\
$\delta_x$ & Masked value for value $x$. \\
$\sigma_x$ & Random mask for value $x$. \\
$\alpha$ & the global MAC key for active security \\
$M$ & An algebraic structure that is either a ring or a finite field. \\
$[x]_M$ & Additive secret sharing of $x$, where $x \in M$. \\
$[[x]]_M$ & Authenticated secret sharing of $x$, where $x \in M$. In passive settings, authentication may be omitted. \\
$\langle x \rangle_M$ & Abstract secret sharing of x, where $x \in M$. \\
\bottomrule
\end{tabularx}
\end{table}

Throughout the paper, a value $x \in M$ is said to be secret-shared if it is distributed among the parties such that no individual party can learn $x$ from its local information alone, where $M$ denotes the underlying algebraic structure. We denote by $\langle x \rangle_M$ an abstract secret sharing of a value $x \in M$, representing that $x$ is distributed among the parties according to the underlying MPC backend. 

Our protocols involve computations over multiple algebraic domains, including arithmetic domains such as $\mathbb{F}_p$ and $\mathbb{Z}_{2^k}$, as well as the binary field $\mathbb{F}_2$ for Boolean operations. When necessary, we explicitly indicate the domain of a shared value using subscripts. The purpose of this abstract sharing semantics is to decouple protocol design from concrete implementation details. Although concrete instantiations of $\langle x \rangle_M$ are introduced later for implementation and performance evaluation, the comparison protocols themselves are defined solely based on the abstract sharing semantics presented here. Table~\ref{tab:notation} summarizes the notation used throughout this paper.

\subsection{Instantiation Examples}
\label{subsec:instantiation}

Concrete instantiations are required to implement the protocols and evaluate their performance. In this work, we instantiate our protocols using SPDZ-style authenticated 
sharing~\cite{keller2020mp}~\cite{feng2022survey}~\cite{cramer2018spd} for arithmetic circuits 
over $M \in \{\mathbb{Z}_{2^k}, \mathbb{F}_p\}$ and TinyOT~\cite{burra2021high} 
for Boolean circuits over $\mathbb{F}_2$~\cite{makri2021rabbit}~\cite{cheng2025mosformer}~\cite{escudero2020improved}. 
We employ optimized representations~\cite{zhang2025md}~\cite{yuan2024md}~\cite{karmakar2024asterisk}: 
$\langle x \rangle_M = (\delta_x, [[\sigma_x]]_M)$ where $\delta_x$ is the 
masked value opened online, $\sigma_x$ is a random mask, and 
$x = \delta_x - \sigma_x \pmod{M}$ (or $x = \delta_x \oplus \sigma_x$ for Boolean).

Table~\ref{tab:instantiation_unified} summarizes the sharing schemes. 
For SPDZ-style sharing, $[[x]]_M = ([x]_M, [m]_M)$ where each party $P_i$ 
holds $([x]_i, [m]_i, [\alpha]_i)$ satisfying MAC relation 
$\sum_{i=1}^{n} [m]_i = (\sum_{i=1}^{n} [x]_i) \cdot (\sum_{i=1}^{n} [\alpha]_i)$ 
over the appropriate domain. For TinyOT, the MAC relation is 
$\sum_{i=1}^{n} [m]_i = (\bigoplus_{i=1}^{n} [x]_i) \cdot (\sum_{i=1}^{n} [\alpha]_i) \in \mathbb{F}_{2^s}$.

\begin{table}[t]
\centering
\caption{Instantiation of Secret-Sharing Schemes}
\label{tab:instantiation_unified}
\small
\setlength{\tabcolsep}{6pt} 
\begin{tabular}{llll}
\toprule
\textbf{Component} & \textbf{$\mathbb{Z}_{2^k}$} (\textbf{SPDZ$_{2^k}$}) & \textbf{$\mathbb{F}_p$} (\textbf{SPDZ}) & \textbf{$\mathbb{F}_2$} (\textbf{TinyOT}) \\
\midrule
Secret $x$ & $\mathbb{Z}_{2^k}$ & $\mathbb{F}_p$ & $\mathbb{F}_2$ \\
Share $[x]$ & $\mathbb{Z}_{2^{k+s}}$ & $\mathbb{F}_p$ & $\mathbb{F}_2$ \\
MAC key $\alpha$ & $\mathbb{Z}_{2^{k+s}}$ & $\mathbb{F}_p$ & $\mathbb{F}_{2^s}$ \\
MAC $[m]$ & $\mathbb{Z}_{2^{k+s}}$ & $\mathbb{F}_p$ & $\mathbb{F}_{2^s}$ \\
Masked $\delta_x$ & $\mathbb{Z}_{2^k}$ & $\mathbb{F}_p$ & $\mathbb{F}_2$ \\
Mask $\sigma_x$ & $\mathbb{Z}_{2^k}$ & $\mathbb{F}_p$ & $\mathbb{F}_2$ \\
\midrule
\multicolumn{4}{l}{\textbf{Sharing:} $[[x]]_M = ([x]_M, [m]_M)$ where $M \in \{\mathbb{Z}_{2^k}, \mathbb{F}_p, \mathbb{F}_2\}$} \\
\multicolumn{4}{l}{\textbf{Optimized:} $\langle x \rangle_M = (\delta_x, [[\sigma_x]]_M)$} \\
\bottomrule
\end{tabular}
\begin{tablenotes}
\footnotesize
\item $s$: statistical security parameter; $k$: plaintext bit length.
\item SPDZ$_{2^k}$ operates over extended ring $\mathbb{Z}_{2^{k+s}}$ for active security.
\end{tablenotes}
\end{table}

\subsection{Rabbit's LTBits Protocol}
\label{protocol:ltbits2-n}

\hyperref[protocol:rabbit-ltbits]{$\Pi_{\text{LTBits}}$} from Rabbit~\cite{makri2021rabbit} provides an efficient construction for comparing a secret bitwise-shared value against a public constant. We adapt this protocol to work with our passive dealer in the preprocessing phase. This adapted protocol forms the foundation for our \hyperref[protocol:ltbits2-n]{$\Pi_{\text{LTBits}_2^n}$} construction in Section~\ref{sec:ltbits2-protocol}.

\begin{protocol}[t]{$\Pi_{\text{LTBits}}$ (Rabbit with dealer version)}
\label{protocol:rabbit-ltbits}

\noindent\textbf{Inputs:} Secret value $x$ shared bitwise, such that parties hold $\langle x_0 \rangle_2, \langle x_1 \rangle_2, \ldots, \langle x_{l-1} \rangle_2$ where $x = \sum\nolimits_{i=0}^{l-1} x_i \cdot 2^i$, and public value $R$ with its bitwise representation $R_0, R_1, \ldots, R_{l-1}$.

\noindent\textbf{Outputs:} The boolean value $\langle z \rangle_2$, where $z = 1\{x < R\}$.

\medskip
\noindent\textbf{Preprocessing Phase:}
\begin{enumerate}
    \item Dealer generates the preprocessing data required by \text{\hyperref[protocol:prefix-and]{$\Pi_{\text{PrefixAND}_2}$}} for $l$-bit inputs.
\end{enumerate}

\noindent\textbf{Online Phase:}
\begin{enumerate}
    \item Parties compute $\langle y_i \rangle_2 = \langle x_i \rangle_2 \oplus R_1 \oplus 1$, where $i \in [0, l-1]$.
    
    \item Parties compute $\langle \vec{q} \rangle_2 = 1 \oplus \text{\hyperref[protocol:prefix-and]{$\Pi_{\text{PrefixAND}_2}$}} (\langle y_0 \rangle_2, \langle y_1 \rangle_2,\\ \ldots, \langle y_{l-1} \rangle_2)$.
    
    \item Parties compute $\langle \vec{w}[i] \rangle_2 = \langle \vec{z}[i] \rangle_2 - \langle \vec{z}[i+1] \rangle_2$,where $i \in [0, l-1]$, $\vec{z}_l = 0$.
    
    \item Result: $\langle z \rangle_2 = \sum\nolimits_{i=0}^{l-1} R_i \cdot \langle \vec{w}[i] \rangle_2$.
\end{enumerate}

\end{protocol}

\section{Comparison with Bitwise-shared Input over $\mathbb{F}_p$}
\label{Sec:ltbits-fp}

This section presents our efficient $n$-party comparison protocol for bitwise-shared inputs over $\mathbb{F}_p$ (\hyperref[protocol:ltbits-fp]{$\Pi_{\text{LTBits}_p}$}). 

\subsection{Protocol Description}
\label{subsec:fp-algorithm}

\begin{protocol}[h]{Protocol 1: $\Pi_{\text{LTBits}_p}$}
\label{protocol:ltbits-fp}

\noindent\textbf{Inputs:} Secret value $x$ shared bitwise, such that parties hold $\langle x_0 \rangle_p, \langle x_1 \rangle_p, \ldots, \langle x_{\ell-1} \rangle_p$ where $x = \sum\nolimits_{i=0}^{\ell-1} x_i \cdot 2^i$, and public value $R$ with its bitwise representation $R_0, R_1, \ldots, R_{\ell-1}$.

\noindent\textbf{Outputs:} Compute the comparison result $\langle b \rangle_p$, where $b = 1\{x < R\}$.

\noindent\textbf{Preprocessing Phase:}
\begin{enumerate}
    \item Dealer computes the polynomial $f^{(\ell)} = \prod\nolimits_{u=1}^{\ell+1} (u-t) = \sum\nolimits_{k=0}^{\ell+1} \alpha_k t^k$ and its coefficient inverse $\gamma = [((\ell+1)!)^{-1}]$, where $\{\alpha_k\}_{k=0}^{\ell+1}, \gamma \in \mathbb{F}_p$.
    
    \item For $i = 0, 1, \ldots, \ell-1$, dealer independently samples $r_i \xleftarrow{\$} \mathbb{F}_p$ and secret-shares $r_i$ among $n$ parties, generating the power sequence $(\langle r_i \rangle_p, \langle r_i^2 \rangle_p, \ldots, \langle r_i^{\ell+1} \rangle_p)$.
    
    \item Dealer distributes $\{\alpha_k\}_{k=0}^{\ell+1}$, $\gamma$ from Step 1 and the $\ell \times (\ell+1)$ secret shares generated in Step 2 to all parties.
\end{enumerate}

\noindent\textbf{Online Phase:}
\begin{enumerate}
    \item Parties compute locally:
    
    \hspace{1em}for $(i=0; i < \ell; i++)$
    
    \hspace{2em}$\langle \vec{u}[i] \rangle_p = \langle \vec{x}[i] \rangle_p - R[i]$
    
    \hspace{2em}$\langle \vec{w}[i] \rangle_p = \langle \vec{x}[i] \rangle_p \oplus R[i]$
    
    \hspace{2em}$\langle \vec{c}[i] \rangle_p = \langle \vec{u}[i] \rangle_p + \left(1 + \sum\nolimits_{k=i+1}^{\ell} \langle \vec{w}[k] \rangle_p\right)$
    
    \item Parties compute $\langle \vec{d}[i] \rangle_p = \langle \vec{c}[i] - r_i \rangle_p$, $i \in [0, \ell-1]$.
    
    \item Parties open $\vec{d}$ to obtain $\vec{d}$ in the clear. \hfill $\triangleright$ One round of communication
    
    \item Parties compute locally:
    
    \hspace{1em}for $(i=0; i < \ell; i++)$
    
    \hspace{2em}for $(k=0; k < \ell+1; k++)$
    
    \hspace{3em}$\langle \vec{z}[i]^k \rangle_p = \sum\nolimits_{j=0}^{k} \binom{k}{j} \langle r_i^j \rangle_p \cdot d_i^{k-j}$
    
    \item Parties compute $\langle \vec{e}[i] \rangle_p = \gamma \cdot \sum\nolimits_{k=0}^{\ell+1} \left\{\alpha_k \cdot \langle \vec{z}[i]^k \rangle_p\right\}$ for $i \in [0, \ell-1]$.
    
    \item Result: $\langle b \rangle_p = \sum\nolimits_{i=0}^{\ell-1} \langle \vec{e}[i] \rangle_p$.
\end{enumerate}

\end{protocol}

Protocol \hyperref[protocol:ltbits-fp]{$\Pi_{\text{LTBits}_p}$} presents the detailed construction of our comparison protocol for bitwise-shared inputs over $\mathbb{F}_p$. Our protocol leverages a passive dealer in the preprocessing phase to shift computational complexity away from the online phase. Specifically, the online phase requires only \textbf{one round of communication} when opening $\vec{d}$ in Step 3, while all other operations are performed locally by the parties.

We note that the computation of $\langle \vec{w}[i] \rangle_p
= \langle \vec{x}[i] \rangle_p \oplus R[i]$ in Step~1 is a purely local
operation, as $R[i]$ is a public bit.  Concretely, this operation corresponds
to a conditional local assignment: if $R[i] = 1$, then
$\langle \vec{w}[i] \rangle_p = \langle \vec{x}[i] \rangle_p$, and if $R[i] = 0$,
then $\langle \vec{w}[i] \rangle_p = 1 - \langle \vec{x}[i] \rangle_p$. 

An important feature of \hyperref[protocol:ltbits-fp]{$\Pi_{\text{LTBits}_p}$} is that the plaintext domain and the secret sharing domain are independent. As long as the plaintext values $x$ and $R$ can be represented in $\ell$ binary bits and $p > \ell + 1$, the protocol executes correctly. This independence implies that \textbf{smaller plaintext domains lead to better performance}: protocols with smaller $\ell$ incur lower computational and communication costs in the online phase, resulting in faster execution.

\section{Comparison with Bitwise-shared Input over $\mathbb{F}_2$}
\label{Sec:ltbits-f2}

In this section, we present our efficient comparison protocol for bitwise-shared inputs over $\mathbb{F}_2$ (\text{\hyperref[protocol:ltbits2-n]{$\Pi_{\text{LTBits}_2^n}$}}). While the previous protocol \hyperref[protocol:ltbits-fp]{$\Pi_{\text{LTBits}_p}$} operates over prime fields $\mathbb{F}_p$, it cannot be directly adapted to the ring $\mathbb{Z}_{2^k}$. The reason is the requirement of multiplying by the coefficient inverse $\gamma = [((l+1)!)^{-1}]$ in the final step. Since not all elements in $\mathbb{Z}_{2^k}$ possess multiplicative inverses, naively migrating \hyperref[protocol:ltbits-fp]{$\Pi_{\text{LTBits}_p}$} to ring-based computation would result in protocol failures.

To address this limitation, we design \text{\hyperref[protocol:ltbits2-n]{$\Pi_{\text{LTBits}_2^n}$}}, which performs computation in the binary field $\mathbb{F}_2$. By leveraging additional $\mathcal{F}_{\text{B2A}}$ conversion, \text{\hyperref[protocol:ltbits2-n]{$\Pi_{\text{LTBits}_2^n}$}} can support execution over both the ring $\mathbb{Z}_{2^k}$ and the field $\mathbb{F}_p$, providing enhanced scalability.

\subsection{Protocol Description}
\label{sec:ltbits2-protocol}

\begin{protocol}[h]{Protocol 2: $\Pi_{\text{AND-}m}$ (TinyOT-style instantiation)}
\label{protocol:and-m}
\noindent\textbf{Inputs:} Secret-shared input bits $\langle x_0 \rangle_2,$ $\langle x_1 \rangle_2,$ $\ldots,$ $\langle x_{m-1} \rangle_2$ in TinyOT-style over $\mathbb{F}_2$.

\noindent\textbf{Outputs:} Secret-shared result $\langle z \rangle_2$ where $z = x_0 \land x_1 \land \cdots \land x_{m-1}$.

\medskip
\noindent\textbf{Preprocessing Phase:}
\begin{enumerate}
    \item Dealer independently samples $r_0, r_1, \ldots, r_{m-1}, \sigma_z $ $\xleftarrow{\$} \mathbb{F}_2$ and generates secret shares $[[\sigma_z]]_2$ and $[[r_i]]_2$ for $i \in \{0, 1, \ldots, m-1\}$, where $\sigma_z$ is the output mask.
    
    \item For each non-empty subset $S \subseteq \{0, 1, \ldots, m-1\}$, dealer computes $\bigwedge_{i \in S} r_i$ and generates secret shares $[[\bigwedge_{i \in S} r_i]]_2$.
    
    \item Dealer computes $\varepsilon_i = r_i \oplus \sigma_{x_i}$ for all $i \in \{0, 1, \ldots,$ $ m-1\}$, where $\sigma_{x_i}$ is the input mask of $\langle x_{i} \rangle_2$.
    
    \item Dealer distributes to all parties:
    \begin{itemize}
        \item Secret shares $[[r_i]]_2$ and $[[\sigma_z ]]_2$ from Step 1
        \item $2^m - 1$ secret shares $[[\bigwedge_{i \in S} r_i]]_2$ from Step 2
        \item Pre-opened values $\{\varepsilon_i\}_{i=0}^{m-1}$
    \end{itemize}
\end{enumerate}

\noindent\textbf{Online Phase:}
\begin{enumerate}
    \item Parties locally compute:
    $$m_i = \delta_{x_i} \oplus \varepsilon_i, \quad i \in \{0, 1, \ldots, m-1\}$$
    
    \item Parties locally compute:
    $$[[\delta_z]]_2 = \sum_{S} \left[ \left(\bigwedge_{i \notin S} m_i\right) \cdot [[\bigwedge_{i \in S} r_i]]_2 \right] \oplus [[\sigma_z]]_2$$
    
    \item Parties open $[[\delta_z]]_2$ to get $\delta_z$, $\langle z \rangle_2 = (\delta_z, [[\sigma_z]]_2)$. \hfill $\triangleright$ One round of communication

\end{enumerate}
\end{protocol}

\begin{protocol}[t]{Protocol 3: $\Pi_{\text{PrefixAND}_n}$}
\label{protocol:prefix-and}
\noindent\textbf{Inputs:} Secret-shared input vector $\langle \vec{x} \rangle_2$ where $x_i \in \mathbb{F}_2$, the size of $\vec{x}$ is $k$ with $i \in [0, k-1]$, and the maximum input size for AND gates is $n$.

\noindent\textbf{Outputs:} Secret-shared output vector $\langle \vec{z} \rangle_2$ where $z_j = \bigwedge_{i=0}^{j} x_i$ for $j \in [0, k-1]$.

\medskip
\noindent\textbf{Algorithm: buildPrefixTree()}

\noindent\textbf{for} $j = 0$ \textbf{to} $\lceil \log_n k \rceil - 1$:

\hspace{1em}\textbf{for} $i = 0$ \textbf{to} $k - 1$: \hfill $\triangleright$ Can be executed in parallel

\hspace{2em}$m = \left\lfloor \frac{i \bmod n^{(j+1)}}{n^j} \right\rfloor + 1$

\hspace{2em}$\text{group\_start} = \left\lfloor \frac{i}{n^{(j+1)}} \right\rfloor \times n^{(j+1)}$

\hspace{2em}$\vec{\text{vec}} = (\text{group\_start} + t \cdot n^j - 1)_{t=1}^{m-1}$

\hspace{2em}$\langle \vec{z}_i \rangle_2 = \text{\hyperref[protocol:and-m]{$\Pi_{\text{AND-}m}$}}(\langle \vec{x}_{\text{vec}} \rangle, \langle x_i \rangle)$

\hspace{1em}Wait for all threads in the inner loop to complete

\medskip
\noindent\textbf{Preprocessing Phase:}
\begin{enumerate}
    \item Execute the preprocessing process required by buildPrefixTree().
\end{enumerate}

\noindent\textbf{Online Phase:}
\begin{enumerate}
    \item Execute the online process required by buildPrefixTree() to obtain $\langle \vec{z} \rangle_2$.
\end{enumerate}
\end{protocol}

The \text{\hyperref[protocol:and-m]{$\Pi_{\text{AND-}m}$}} and \text{\hyperref[protocol:prefix-and]{$\Pi_{\text{PrefixAND}_n}$}} protocols provide the specific construction for $m$-input AND gates and PrefixAND circuits composed of AND gates with maximum input size $n$. We note that \text{\hyperref[protocol:and-m]{$\Pi_{\text{AND-}m}$}} presents a TinyOT-style instantiation of the $m$-input AND gate, chosen to demonstrate an engineering optimization that achieves one-round online complexity through preprocessing assistance. However, we emphasize that constant-round realizations of $m$-input AND gates exist across all mainstream MPC backends, including Shamir secret sharing~\cite{reistad2009multiparty}, Replicated secret sharing~\cite{mohassel2018aby3}, and SPDZ-style secret sharing~\cite{patra2021aby2}. 

\begin{figure}[!t]
\centering
\includegraphics[width=0.48\textwidth]{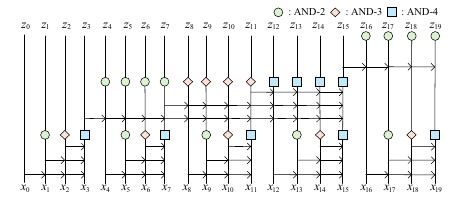}
\caption{Example of PrefixAND circuit with input length $k=20$ and maximum branching factor $n=4$.}
\label{fig:prefix-and-example}
\end{figure}

Figure~\ref{fig:prefix-and-example} illustrates an example of a PrefixAND circuit with input length $k=20$ and maximum branching factor $n=4$. Using the following formula, we can transform a PrefixAND circuit into a PrefixOR circuit:
\[z_i = \bigvee_{j=0}^{i} x_j = 1 \oplus \bigwedge_{j=0}^{i} (1 \oplus x_j)\] 
Following the construction methodology of Rabbit's LTBits protocol~\cite{makri2021rabbit}, we can construct our \text{\hyperref[protocol:ltbits2-n]{$\Pi_{\text{LTBits}_2^n}$}} protocol. The detailed construction is presented in Section~\ref{protocol:ltbits2-n}. The communication rounds of \text{\hyperref[protocol:ltbits2-n]{$\Pi_{\text{LTBits}_2^n}$}} depend on the maximum AND gate's input size $n$ and the bit length $k$ of plaintext $x$, specifically $O(\log_n k)$ rounds with $O(k \cdot \log_n k)$ communication cost.

\section{Extension to MSB Protocol}
\label{Sec:msb}

While the comparison protocols \text{\hyperref[protocol:ltbits-fp]{$\Pi_{\text{LTBits}_p}$}} and \text{\hyperref[protocol:ltbits2-n]{$\Pi_{\text{LTBits}_2^n}$}} presented in previous sections provide efficient solutions for determining whether $x < R$ (where $R$ is a public value), they require the input $\langle x \rangle$ to be in bitwise-shared form. This design lacks the generality needed for arithmetic circuits where values are typically shared in their native arithmetic representation. The Most Significant Bit (MSB) extraction protocol serves as a crucial building block that bridges this gap and enables a rich family of comparison functionalities.

By extending our LTBits protocols to MSB extraction, we can derive a comprehensive suite of comparison protocols that operate on arithmetically shared inputs, including:
\begin{itemize}
    \item $\Pi_{\text{LTS}}$~\cite{makri2021rabbit}: Less-than comparison for secret-shared values, determining whether $x < y$ where both $x$ and $y$ are given in secret-shared form.
    \item $\Pi_{\text{Select}}$~\cite{cheng2025mosformer}: $\text{Select}(x, y) = 1\{x \geq 0\} \cdot y$. The protocol $\Pi_{\text{Select}}$ outputs $\langle y \rangle$ if $x \geq 0$, and $\langle 0 \rangle$ otherwise.
    \item $\Pi_{\text{Max}}$~\cite{cheng2025mosformer}: Maximum value computation between secret-shared inputs.
    \item $\Pi_{\text{ReLU}}$~\cite{zhang2025md}~\cite{makri2021rabbit}: Rectified Linear Unit activation function where $\text{ReLU}(x) = \max(0, x)$.
    \item $\Pi_{\text{EQZ}}$~\cite{damgaard2019new}: Equality-to-zero test for secret-shared values.
\end{itemize}
These derived protocols provide diverse and flexible comparison primitives that are essential for complex secure computation tasks, such as privacy-preserving machine learning, secure auctions.

This section is organized as follows: First, we present the methodology for extending \text{\hyperref[protocol:ltbits-fp]{$\Pi_{\text{LTBits}_p}$}} and \text{\hyperref[protocol:ltbits2-n]{$\Pi_{\text{LTBits}_2^n}$}} to MSB extraction in Section~\ref{sec:extending-to-msb}. Then, we discuss the perfect security of \text{\hyperref[protocol:msb-p]{$\Pi_{\text{MSB}_p}$}} and \text{\hyperref[protocol:msb-2k]{$\Pi_{\text{MSB}_{2^k}}$}}. In Section I of supplementary file, we analyzes an incorrect MSB construction from prior work.

\subsection{Extending $\Pi_{\text{LTBits}_p}$ and $\Pi_{\text{LTBits}_2^n}$ to MSB}
\label{sec:extending-to-msb}

We consider the MSB construction from two cases based on the underlying arithmetic secret sharing domain.

\textbf{Case 1: MSB over $\mathbb{F}_p$.} 
In the first case, arithmetic secret sharing is performed over the prime field $\mathbb{F}_p$, meaning that both the plaintext domain and the arithmetic secret sharing domain are $\mathbb{F}_p$. The MSB operation is defined as $\text{MSB}(x) = 1\{x \geq \lfloor p/2 \rfloor\}$. For this scenario, we adopt the Less-Than-Constant (LTC) construction approach from Rabbit~\cite{makri2021rabbit}, where the constant is set to $\lfloor p/2 \rfloor$. 

Our construction differs from Rabbit in two aspects: (1) Since \text{\hyperref[protocol:ltbits-fp]{$\Pi_{\text{LTBits}_p}$}} protocol operates directly in $\mathbb{F}_p$, we do not require the final $\mathcal{F}_{\text{B2A}}$ conversion step to transform the result from $\langle z \rangle_2$ to $\langle z \rangle_p$. (2) In contrast to Rabbit’s~\cite{makri2021rabbit} $\mathbb{F}_p$ instantiation, which relies on edaBits~\cite{escudero2020improved} and thus only achieves statistical security due to potential wrap-around inconsistencies, the preprocessing phase of our \hyperref[protocol:msb-p]{$\Pi_{\text{MSB}_p}$} protocol is carried out by a passive dealer that samples $r \in [0,p-1]$ and provides a \textit{perfectly consistent} arithmetic/bit decomposition of $r$. As a result, our \hyperref[protocol:msb-p]{$\Pi_{\text{MSB}_p}$} does not incur statistical errors and achieves perfect security~\cite{evans2018pragmatic}. \hyperref[protocol:msb-p]{$\Pi_{\text{MSB}_p}$} provides the detailed construction of MSB over $\mathbb{F}_p$. 

Note that constructing the MSB protocol over $\mathbb{F}_p$ using \text{\hyperref[protocol:ltbits2-n]{$\Pi_{\text{LTBits}_2^n}$}} follows an identical procedure, with the only additional step being a $\mathcal{F}_{\text{B2A}}$ conversion. We omit the details here for brevity.

\textbf{Case 2: MSB over $\mathbb{Z}_{2^k}$.}
In the second case, arithmetic secret sharing is performed over the ring $\mathbb{Z}_{2^k}$, meaning the plaintext domain is $\mathbb{Z}_{2^k}$. Under the passive security model, arithmetic secret shares are in $\mathbb{Z}_{2^k}$, while under the active security model, they are in $\mathbb{Z}_{2^{k+s}}$~\cite{cramer2018spd}. For this scenario, our $\Pi_{\text{MSB}_{2^k}}$ protocol draws inspiration from~\cite{fu2024private,gupta2024sigma,cheng2025mosformer} using the following formula:
\[\text{MSB}(x) = \text{MSB}(\tilde{x}) \oplus \text{MSB}(2^k - r) \oplus 1\{\tilde{y_0} > 2^{k-1} - \tilde{y_1} - 1\}\]

In our construction, this translates to:
\[\text{MSB}(x) = \text{MSB}(\tilde{x}) \oplus \text{MSB}(2^k - r) \oplus \Pi_{\text{LTBits}_2^n}(\langle 2^{k-1} - \tilde{y_1} - 1 \rangle, \tilde{y_0})\]

where $r$ is a random mask in $\mathbb{Z}_{2^k}$, $\tilde{x} = x + r \bmod 2^k$, $\tilde{y_0} = \tilde{x} \bmod 2^{k-1}$, and $\tilde{y_1} = (2^k - r) \bmod 2^{k-1}$.

Our \text{\hyperref[protocol:ltbits2-n]{$\Pi_{\text{LTBits}_2^n}$}} protocol provides unique advantages for this construction. First, $\text{MSB}(\tilde{x})$ can be obtained by opening in the online phase. Second, $\text{MSB}(2^k - r)$ is generated as $\langle \text{MSB}(2^k - r) \rangle$ by the dealer during the preprocessing phase. The introduction of the dealer eliminates the need for an additional MSB protocol invocation in the preprocessing phase. Finally, a $\mathcal{F}_{\text{B2A}}$ converts the result from $\langle z \rangle_2$ to $\langle z \rangle_{2^k}$. In \text{\hyperref[protocol:msb-2k]{$\Pi_{\text{MSB}_{2^k}}$}}, $\tilde{x}$ revealed during the online phase are masked by uniformly random elements over $\mathbb{Z}_{2^k}$, ensuring information-theoretic privacy. Meanwhile dealer generates $\langle 2^{k-1} - \hat{y}_1 - 1 \rangle_2$ in bitwise representation, the bit-decomposition of $2^{k-1} - \hat{y}_1 - 1$ is perfectly consistent. As a result, our \text{\hyperref[protocol:ltbits2-n]{$\Pi_{\text{LTBits}_2^n}$}} does not incur statistical errors and achieves perfect security~\cite{evans2018pragmatic}.
\text{\hyperref[protocol:msb-2k]{$\Pi_{\text{MSB}_{2^k}}$}}
 provides the detailed construction of MSB over $\mathbb{Z}_{2^k}$. Our construction ensures that the online runtime of \text{\hyperref[protocol:msb-2k]{$\Pi_{\text{MSB}_{2^k}}$}} is close to that of \text{\hyperref[protocol:ltbits2-n]{$\Pi_{\text{LTBits}_2^n}$}}.

\begin{protocol}[t]{Protocol 4: $\Pi_{\text{MSB}_p}$ (with \text{$\Pi_{\text{LTBits}_p}$})}
\label{protocol:msb-p}
\noindent\textbf{Inputs:} Secret-shared value $\langle x \rangle_p$ over $\mathbb{F}_p$, and public value $\lfloor p/2 \rfloor$.

\noindent\textbf{Outputs:} The Boolean value $\langle z \rangle_p = \langle 1\{x \geq \lfloor p/2 \rfloor\} \rangle_p$.

\medskip
\noindent\textbf{Preprocessing Phase:}
\begin{enumerate}
    \item Dealer independently samples $r \xleftarrow{\$} \mathbb{F}_p$, computes its binary expansion $r = \sum_{i=0}^{m-1} r_i \cdot 2^i$, where $m$ denotes the bit length of prime $p$. Dealer generates secret shares $\langle r \rangle_p, \langle r_0 \rangle_p, \langle r_1 \rangle_p, \ldots, \langle r_{m-1} \rangle_p$.
    
    \item Dealer generates double copies of the preprocessing data required by \text{\hyperref[protocol:ltbits-fp]{$\Pi_{\text{LTBits}_p}$}} for $m$-bit inputs.
    
    \item Dealer distributes the generated shares $\langle r \rangle_p, \langle r_0 \rangle_p,$ $ \ldots, \langle r_{m-1} \rangle_p$ and the \text{\hyperref[protocol:ltbits-fp]{$\Pi_{\text{LTBits}_p}$}} preprocessing data to all parties.
\end{enumerate}

\noindent\textbf{Online Phase:}
\begin{enumerate}
    \item Parties compute the values:
    $\langle a \rangle_p = \langle x + r \rangle_p \quad $\text{and} $\quad \langle b \rangle_p = \langle x + r + \lceil p/2 \rceil \rangle_p$.
    
    \item Parties open the values $a$ and $b$. 
    
    \item Parties compute: \hfill $\triangleright$ Can be executed in parallel
    
    \hspace{1em}$\langle w_1 \rangle_p = \text{\hyperref[protocol:ltbits-fp]{$\Pi_{\text{LTBits}_p}$}}(a, \langle r_0 \rangle_p, \ldots, \langle r_{m-1} \rangle_p)$
    
    \hspace{1em}$\langle w_2 \rangle_p = \text{\hyperref[protocol:ltbits-fp]{$\Pi_{\text{LTBits}_p}$}}(b, \langle r_0 \rangle_p, \ldots, \langle r_{m-1} \rangle_p)$
    
    \hspace{1em}$w_3 = 1\{b < \lceil p/2 \rceil\}$
    
    \item Result: $\langle z \rangle_p = \langle w_1 \rangle_p - \langle w_2 \rangle_p + w_3$
\end{enumerate}
\end{protocol}

\begin{protocol}[t]{Protocol 5: $\Pi_{\text{MSB}_{2^k}}$ (with \text{$\Pi_{\text{LTBits}_2^n}$})}
\label{protocol:msb-2k}
\noindent\textbf{Inputs:} Secret-shared value $\langle x \rangle_{2^k}$ over $\mathbb{Z}_{2^k}$, and AND gate's maximum branching factor $n$.

\noindent\textbf{Outputs:} The Boolean value $\langle z \rangle_{2^k} = \langle 1\{x < 0\} \rangle_{2^k}$.

\medskip
\noindent\textbf{Preprocessing Phase:}
\begin{enumerate}
    \item Dealer independently samples $r \xleftarrow{\$} \mathbb{Z}_{2^k}$ and computes $\hat{y}_1 = (2^k - r) \bmod 2^{k-1}$.
    
    \item Dealer generates $\langle \text{MSB}(2^k - r) \rangle_2$ and generates $\langle 2^{k-1} - \hat{y}_1 - 1 \rangle_2$ in bitwise representation.
    
    \item Dealer generates the preprocessing data required by \text{\hyperref[protocol:ltbits2-n]{$\Pi_{\text{LTBits}_2^n}$}} for $k$-bit inputs.
    
    \item Dealer distributes to all parties: $\langle r \rangle_{2^k}$, $\langle \text{MSB}(2^k - r) \rangle_2$, $\langle 2^{k-1} - \hat{y}_1 - 1 \rangle_2$ in bitwise representation, and the preprocessing data from Step 3.
\end{enumerate}

\noindent\textbf{Online Phase:}
\begin{enumerate}
    \item Parties compute $\langle \hat{x} \rangle_{2^k} = \langle x \rangle_{2^k} + \langle r \rangle_{2^k}$ and open $\hat{x}$. 
    
    \item Parties compute $\hat{y}_0 = \hat{x} \bmod 2^{k-1}$, then compute:
    $$\langle w \rangle_2 = \text{\hyperref[protocol:ltbits2-n]{$\Pi_{\text{LTBits}_2^n}$}}(\langle 2^{k-1} - \hat{y}_1 - 1 \rangle_2^{\text{bitwise}}, \hat{y}_0)$$
    
    \item Parties compute:
    $$\langle z \rangle_2 = \text{MSB}(\hat{x}) \oplus \langle \text{MSB}(2^k - r) \rangle_2 \oplus \langle w \rangle_2$$
    
    \item Parties invoke $\mathcal{F}_{\text{B2A}}$ to get $\langle z \rangle_{2^k}$ from $\langle z \rangle_2$.
    
\end{enumerate}
\end{protocol}

\subsection{Discussion of Perfect Security}
\label{sec:perfect-security}

In this section, we clarify the notion of \emph{perfect security} and \emph{statistical security}. Then we explain why \hyperref[protocol:msb-p]{$\Pi_{\text{MSB}_p}$} and \text{\hyperref[protocol:msb-2k]{$\Pi_{\text{MSB}_{2^k}}$}} achieve perfect security \emph{at the protocol
level}.  We emphasize that the discussion here is restricted to
the abstract protocol-level description and does not take into account any
concrete backend realization.

\paragraph{Perfect Security vs.\ Statistical Security. (Complete Definition in A.3 of \cite{crepeau2002secure})}
In the context of MPC, a protocol is said to achieve perfect (information-theoretic) security
if the view of any adversary is \emph{exactly independent} of the honest parties'
private inputs, except for what can be inferred from the prescribed output.
Equivalently, the real execution and the ideal execution are distributed
identically, without introducing any error probability or security parameter.

In contrast, a protocol is statistically secure if the adversary's view is
only \emph{approximately independent} of the private inputs.  In this case,
the statistical distance between the real and ideal executions is bounded by a
negligible function in a security parameter~$s$, and the protocol may fail with
probability at most $2^{-s}$.  Such statistical guarantees commonly arise from
consistency checks, authentication mechanisms, or probabilistic preprocessing
procedures.

\paragraph{Perfect Security of \hyperref[protocol:msb-p]{$\Pi_{\text{MSB}_p}$}.}
Recall that in \hyperref[protocol:msb-p]{$\Pi_{\text{MSB}_p}$}, the parties first open masked values of $x + r$ over $\mathbb{F}_p$, where $r$ is a uniformly
random field element generated during preprocessing and used exactly once.
Since $r$ is information-theoretically uniform, the opened value $x + r$ is
itself uniformly distributed over $\mathbb{F}_p$ and statistically independent
of the secret input $x$.

Unlike existing $\mathbb{F}_p$-based MSB constructions such as Rabbit~\cite{makri2021rabbit}, which rely on edaBits~\cite{escudero2020improved} and may incur statistical errors due to arithmetic/bit decomposition
inconsistencies, the preprocessing phase of
\hyperref[protocol:msb-p]{$\Pi_{\text{MSB}_p}$} is carried out by a passive dealer that provides a perfectly consistent arithmetic and bitwise decomposition of the random mask $r$.  As a result, no statistical deviation is introduced.

All subsequent computations operate either on public values or on secret-shared
randomness provided by the preprocessing phase. Therefore, the adversary’s view is identically distributed for all possible inputs $x$,
implying that \hyperref[protocol:msb-p]{$\Pi_{\text{MSB}_p}$} achieves perfect security at the protocol level.

\paragraph{Perfect Security of \text{\hyperref[protocol:msb-2k]{$\Pi_{\text{MSB}_{2^k}}$}}.}
A similar argument applies to \text{\hyperref[protocol:msb-2k]{$\Pi_{\text{MSB}_{2^k}}$}}.  In this protocol, the
parties open the masked value $\hat{x} = x + r \bmod 2^k$, where $r$ is uniformly
random over $\mathbb{Z}_{2^k}$.  By the one-time masking property, $\hat{x}$ is
uniformly distributed and independent of $x$, and therefore reveals no
information about the secret input.

The remaining steps only involve comparisons between public values derived from
$\hat{x}$ and secret-shared values derived from $r$, together with fixed public
constants.  Since all secret-shared inputs to these subprotocols originate from
perfectly random preprocessing material and are independent of $x$, the entire
execution transcript is information-theoretically independent of the private
input.  Consequently, \text{\hyperref[protocol:msb-2k]{$\Pi_{\text{MSB}_{2^k}}$}} also achieves perfect security at the protocol level.

\section{Evaluation Performance}
\label{sec:evaluation}

In this section, we evaluate the performance of our proposed comparison protocols against Rabbit~\cite{makri2021rabbit}, which represents the state-of-the-art in MPC-based comparison protocols. We organize the evaluation as follows: Section~\ref{subsec:experimental_setup} describes the experimental setup including hardware configuration, network settings, and implementation details. Section~\ref{subsec:eval_msb_p} presents the performance evaluation of \hyperref[protocol:msb-p]{$\Pi_{\text{MSB}_p}$} over $\mathbb{F}_p$. Section~\ref{subsec:eval_msb_2k} evaluates \hyperref[protocol:msb-2k]{$\Pi_{\text{MSB}_{2^k}}$} over $\mathbb{Z}_{2^k}$.

\subsection{Experimental Setup}
\label{subsec:experimental_setup}

\paragraph{Hardware and software configuration.}
All experiments are conducted on a machine running Ubuntu 20.04.5 LTS equipped with a 2.2GHz AMD Ryzen Threadripper 3970X 32-Core Processor (64 logical cores) and 251GB of RAM. We implement all protocols in \texttt{C++20} for efficient cryptographic computation.

\paragraph{Implementation framework.}
Since the MP-SPDZ library~\cite{keller2020mp} does not provide the sophisticated preprocessing functionality required by our dealer-assisted protocols, we follow the implementation paradigm established by MD-ML~\cite{yuan2024md}, which similarly relies on a passive dealer for preprocessing. Our codebase adopts the same modular architecture to facilitate dealer-based preprocessing while maintaining compatibility with standard MPC backend interfaces.

\paragraph{Network configuration.}
To evaluate protocol performance under realistic network conditions, we employ the Linux \texttt{Traffic Control} utility to simulate two representative network settings:
\begin{itemize}
    \item \textbf{WAN (Wide Area Network):} 100ms network delay, 1\% packet loss rate, and 100Mbps bandwidth.
    \item \textbf{LAN (Local Area Network):} 1ms network delay, 0.01\% packet loss rate, and 10Gbps bandwidth.
\end{itemize}

\paragraph{MPC backend instantiation.}
As described in Section~\ref{subsec:instantiation}, we instantiate our protocols using two representative MPC backends. Both instantiations support dishonest-majority adversaries and enforce active security through MAC-based verification during the opening phase.

\subsection{Performance of $\Pi_{\text{MSB}_p}$}
\label{subsec:eval_msb_p}

We evaluate the performance of \hyperref[protocol:msb-p]{$\Pi_{\text{MSB}_p}$} across different prime field sizes, security models, and network configurations. We compare our protocol against Rabbit's MSB protocol over $\mathbb{F}_p$ under identical experimental conditions. The evaluation considers three representative prime choices: $p = 2^{16} - 15$ (16-bit plaintext domain), $p = 2^{31} - 1$ (31-bit plaintext domain), and $p = 2^{61} - 1$ (61-bit plaintext domain). For each prime, we measure both runtime and communication cost under passive and active security models in LAN and WAN settings.

\begin{figure*}[!t]
\centering
\includegraphics[width=0.95\textwidth]{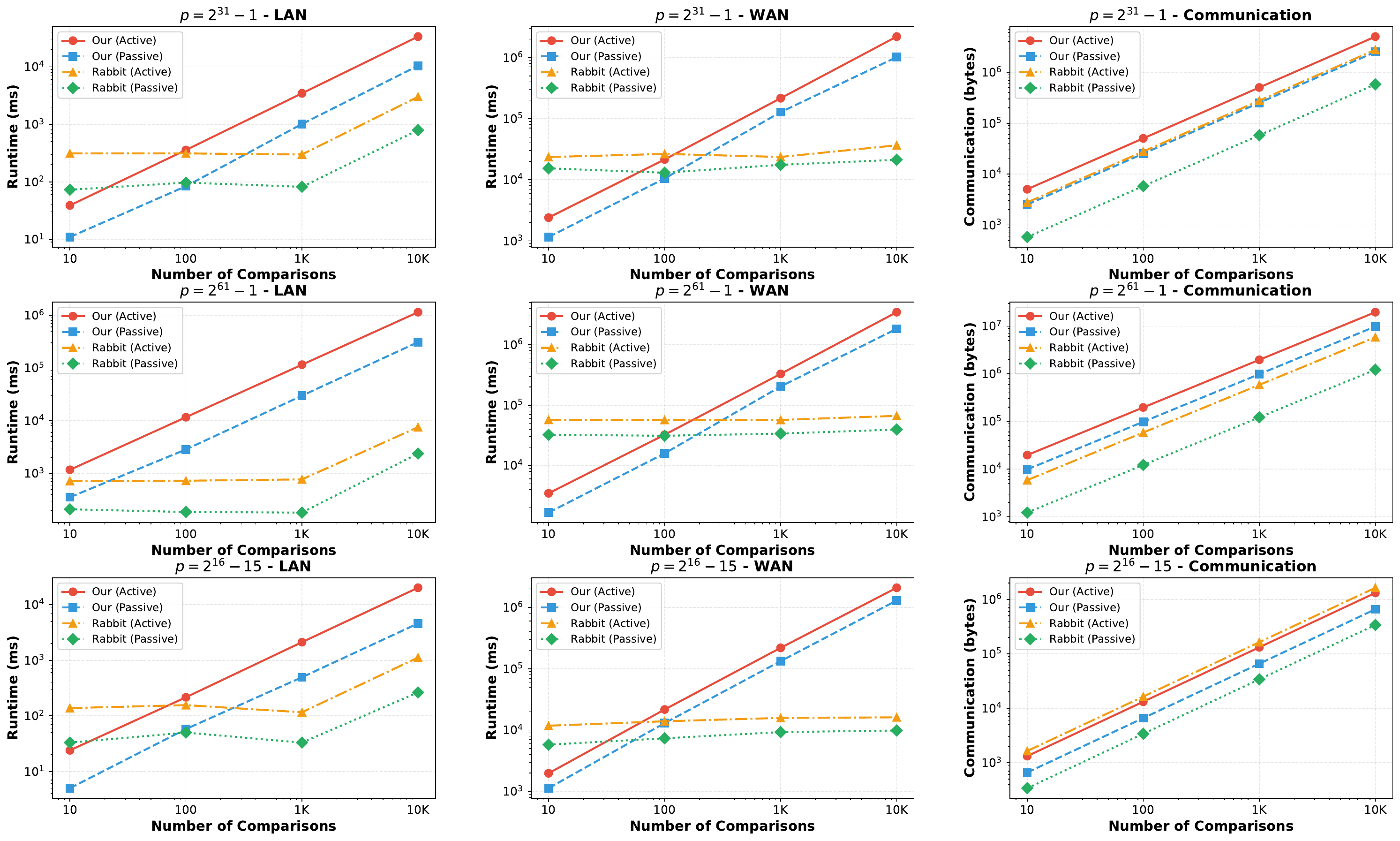}
\caption{Performance comparison between \hyperref[protocol:msb-p]{$\Pi_{\text{MSB}_p}$} and Rabbit's MSB protocol over $\mathbb{F}_p$ under different prime field sizes, security models, and network configurations. Each row represents a different prime $p$, while columns show LAN runtime, WAN runtime, and communication cost respectively.}
\label{fig:msb_p_performance}
\end{figure*}

Figure~\ref{fig:msb_p_performance} presents the comprehensive performance comparison across 10, 100, 1,000, and 10,000 comparison operations. The results are organized in a 3$\times$3 grid, where each row corresponds to a specific prime field size and each column represents a different performance metric: LAN runtime, WAN runtime, and communication cost.

\paragraph{Performance analysis and recommended usage scenarios.}
The experimental results reveal that \hyperref[protocol:msb-p]{$\Pi_{\text{MSB}_p}$} demonstrates significant performance advantages in specific operational regimes. Our protocol achieves substantial speedups over Rabbit when the number of comparison operations is relatively small (typically fewer than 100 comparisons) and the prime field size is modest. This advantage is particularly pronounced in WAN settings.

For the smallest tested prime ($p = 2^{16} - 15$), our protocol achieves $5.75\times$ to $6.6\times$ speedup in LAN settings and $5.11\times$ to $5.96\times$ speedup in WAN settings when performing 10 comparisons. Under the medium-sized prime ($p = 2^{31} - 1$), the speedup increases to $6.6\times$ to $8.0\times$ in LAN and reaches $9.8\times$ to $13.3\times$ in WAN. Notably, for the largest prime ($p = 2^{61} - 1$), while Rabbit maintains superior performance in LAN environments, our protocol still achieves remarkable speedups of $16.5\times$ to $19.4\times$ in WAN settings for small-scale comparisons.

The performance characteristics stem from two key factors. First, our protocol's reduced round complexity provides substantial benefits when the number of comparisons is small, as the per-round network latency overhead dominates in WAN environments. Second, smaller prime fields reduce the local computational burden of polynomial evaluation in our construction, making the constant factor overhead more acceptable. As the number of comparisons increases beyond 100, Rabbit's optimized circuit construction and better asymptotic complexity begin to dominate, particularly in LAN settings where round complexity has less impact.

Table~\ref{tab:msb_p_recommendations} summarizes the recommended operational regimes for \hyperref[protocol:msb-p]{$\Pi_{\text{MSB}_p}$} across different configurations. \hyperref[protocol:msb-p]{$\Pi_{\text{MSB}_p}$} is more effective for applications requiring small batches of comparisons over modest-sized fields, particularly in network-constrained environments.

\begin{table}[t]
\centering
\caption{Recommended Usage for $\Pi_{\text{MSB}_p}$ and Maximum Speedup}
\label{tab:msb_p_recommendations}
\small
\begin{tabular}{llcc}
\toprule
\textbf{Prime} & \textbf{Security} & \textbf{Recommended} & \textbf{Max Speedup} \\
               &                   & \textbf{Range$^\dagger$}      & \textbf{(LAN/WAN)} \\
\midrule
$2^{16} - 15$ & Active   & $< 100$   & $5.8\times$ / $6.0\times$ \\
              & Passive  & $< 100$   & $6.6\times$ / $5.1\times$ \\
\midrule
$2^{31} - 1$  & Active   & $< 100$ / $< 1000$  & $8.0\times$ / $9.8\times$ \\
              & Passive  & $< 1000$ / $< 1000$  & $6.6\times$ / $13.3\times$ \\
\midrule
$2^{61} - 1$  & Active   & -- / $< 100$   & -- / $16.5\times$ \\
              & Passive  & -- / $< 100$   & -- / $19.4\times$ \\
\bottomrule
\end{tabular}
\begin{tablenotes}
\footnotesize
\item $^\dagger$: Range favoring our protocol (LAN / WAN format).
\item "--" indicates Rabbit is faster across all tested counts.
\end{tablenotes}
\end{table}

\subsection{Performance of $\Pi_{\text{MSB}_{2^k}}$}
\label{subsec:eval_msb_2k}

We evaluate the performance of \text{\hyperref[protocol:msb-2k]{$\Pi_{\text{MSB}_{2^k}}$}} across different ring sizes and AND gate branching factors. The evaluation considers two representative configurations: $k=s=32$ and $k=s=64$, $k$ represents the size of plaintext domain, $s$  represents statistical security parameter. For each configuration, we test our protocol for 10000 comparison operations with varying maximum AND gate input sizes $n \in [2,10]$, measuring both runtime and communication cost under passive and active security models in LAN and WAN settings.

\begin{figure*}[!t]
\centering
\includegraphics[width=\textwidth]{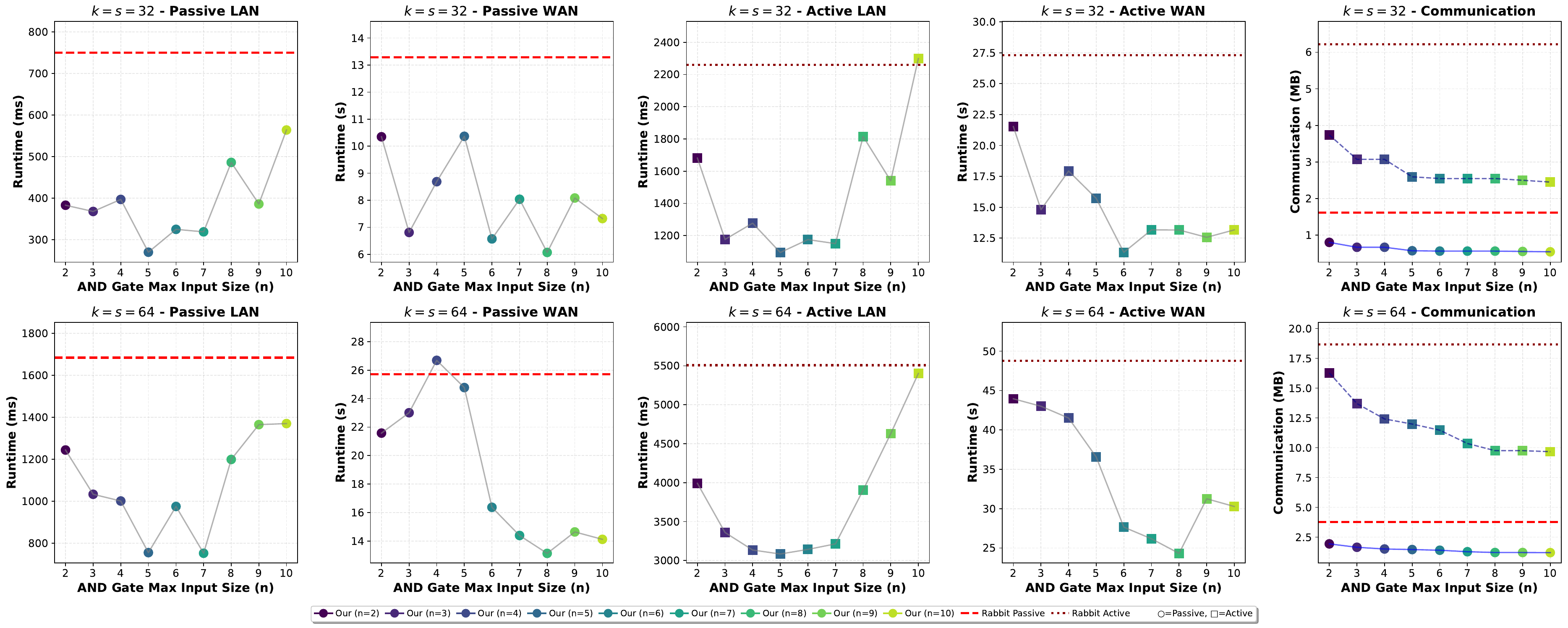}
\caption{Performance comparison between \text{\hyperref[protocol:msb-2k]{$\Pi_{\text{MSB}_{2^k}}$}} and Rabbit's MSB protocol over $\mathbb{Z}_{2^k}$ for 10000 comparison operations. Each subplot shows our protocol's performance across different AND gate branching factors ($n=2$ to $10$) versus Rabbit's baseline (restricted in $n=2$). The first row shows results for $k=s=32$, and the second row for $k=s=64$.}
\label{fig:msb_2k_performance}
\end{figure*}

Figure~\ref{fig:msb_2k_performance} presents the comprehensive performance comparison for 10000 comparison operations across different branching factors. The results are organized in a 2$\times$5 grid. The first row shows results for $k=s=32$, while the second row presents $k=s=64$. Each row displays five performance metrics: passive LAN runtime, passive WAN runtime, active LAN runtime, active WAN runtime, and communication cost.

\paragraph{Performance analysis and optimal branching factor selection.}
The experimental results demonstrate that the optimal choice of AND gate branching factor $n$ varies significantly across different configurations. Table~\ref{tab:msb_2k_optimal_n} summarizes the optimal $n$ and corresponding performance comparisons with Rabbit for 10,000 comparison operations.

For the $k=s=32$ configuration, \text{\hyperref[protocol:msb-2k]{$\Pi_{\text{MSB}_{2^k}}$}} achieves substantial speedups across all scenarios. Under passive security, the optimal branching factor is $n=5$ for LAN environments, yielding 270ms runtime compared to Rabbit's 750ms---a $2.78\times$ speedup. In WAN settings, larger branching factors become preferable due to the dominance of network latency: $n=8$ provides 6.07 seconds runtime versus Rabbit's 13.3 seconds, achieving a $2.19\times$ speedup. Under active security, the pattern holds with optimal $n$ values of 5 for LAN and 6 for WAN, delivering speedups of $2.06\times$ and $2.40\times$ respectively.

For the larger $k=s=64$ configuration, the benefits of increased branching become more pronounced for runtime reduction. Passive security achieves optimal performance at $n=7$ in LAN (752ms vs 1,684ms, $2.24\times$ speedup) and $n=8$ in WAN (13.1s vs 25.7s, $1.96\times$ speedup). Active security shows similar trends with optimal values at $n=5$ for LAN ($1.79\times$ speedup) and $n=7$ for WAN ($1.86\times$ speedup).

The results reveal several key insights. First, WAN environments consistently benefit from larger branching factors ($n \geq 6$) compared to LAN settings, as the reduction in communication rounds outweighs the increased per-round overhead when network latency is high. Second, optimal $n$ values tend to increase with larger plaintext domains ($k=s=64$ favors $n=7$-$8$ vs $k=s=32$ favoring $n=5$-$6$), reflecting the greater benefit of reducing rounds when processing more bits. Third, passive security generally permits slightly larger optimal $n$ values than active security due to lower per-round overhead. Across all configurations, our protocol \text{\hyperref[protocol:msb-2k]{$\Pi_{\text{MSB}_{2^k}}$}} achieves $1.79\times$ to $2.78\times$ speedup over Rabbit, demonstrating the effectiveness of multi-input AND gate optimization.

\begin{table}[t]
\centering
\caption{Optimal Factor and Performance Comparison for $\Pi_{\text{MSB}_{2^k}}$}
\label{tab:msb_2k_optimal_n}
\small
\begin{tabular}{lccccc}
\toprule
\textbf{Config} & \textbf{Network} & \textbf{Opt.} & \textbf{Our} & \textbf{Rabbit} & \textbf{Speed-} \\
                         &                  & \textbf{$n$}  & \textbf{Time} & \textbf{Time} & \textbf{up} \\
\midrule
\multirow{2}{*}{$32$ Passive}
& LAN & 5 & 270ms & 750ms & $2.78\times$ \\
& WAN & 8 & 6.1s  & 13.3s & $2.19\times$ \\
\multirow{2}{*}{$32$ Active}
& LAN & 5 & 1.1s  & 2.3s  & $2.06\times$ \\
& WAN & 6 & 11.4s & 27.3s & $2.40\times$ \\
\midrule
\multirow{2}{*}{$64$ Passive}
& LAN & 7 & 752ms & 1.7s  & $2.24\times$ \\
& WAN & 8 & 13.1s & 25.7s & $1.96\times$ \\
\multirow{2}{*}{$64$ Active}
& LAN & 5 & 3.1s  & 5.5s  & $1.79\times$ \\
& WAN & 7 & 26.2s & 48.8s & $1.86\times$ \\
\bottomrule
\end{tabular}
\begin{tablenotes}
\footnotesize
\item Config: 32/64 denotes $k=s$ values. 
\end{tablenotes}
\end{table}

\section{Conclusion}
\label{Sec:conclusion}

In this work, we presented the first dealer-assisted LTBits and MSB extraction 
protocols over both $\mathbb{F}_p$ and $\mathbb{Z}_{2^k}$, achieving perfect 
security at the protocol level while supporting general $n$-party computation 
under diverse adversary models. Our polynomial-based construction for $\mathbb{F}_p$ 
achieves constant-round online complexity, while our multi-input AND gate 
optimization for $\mathbb{Z}_{2^k}$ reduces round complexity to $O(\log_n k)$ 
with tunable branching factors. Experimental evaluations demonstrate substantial 
speedups of $1.79\times$ to $19.4\times$ over state-of-the-art frameworks.

Our protocols provide a strong foundation for privacy-preserving machine learning, 
where comparison operations are fundamental in neural network inference (ReLU, 
max pooling) and training (gradient clipping). Future work includes integrating 
our protocols into end-to-end PPML frameworks and extending our techniques to 
related primitives such as secure division and truncation.


\bibliographystyle{IEEEtran}
\bibliography{ref}

\vfill

\end{document}